\documentclass[prl,twocolumn,showpacs]{revtex4}
\usepackage{amsmath,amssymb} 
\usepackage{bm}
\usepackage{hyperref}
\usepackage{subfigure}
\usepackage{graphicx}
\pdfoutput=1

\begin{document}

\title{Atoms in a radiofrequency-dressed optical lattice}
\author{N.~Lundblad} \email{nathan.lundblad@nist.gov}
\affiliation{Joint Quantum Institute, National Institute of Standards and Technology and University of Maryland, 
Gaithersburg, Maryland 20899, USA}

\author{P. J.~Lee}
\affiliation{Joint Quantum Institute, National Institute of Standards and Technology and University of Maryland, 
Gaithersburg, Maryland 20899, USA}

\author{I. B.~Spielman}
\affiliation{Joint Quantum Institute, National Institute of Standards and Technology and University of Maryland, 
Gaithersburg, Maryland 20899, USA}

\author{B. L.~Brown}
\affiliation{Joint Quantum Institute, National Institute of Standards and Technology and University of Maryland, 
Gaithersburg, Maryland 20899, USA}

\author{W. D.~Phillips}
\affiliation{Joint Quantum Institute, National Institute of Standards and Technology and University of Maryland, 
Gaithersburg, Maryland 20899, USA}

 \author{J. V.~Porto}
\affiliation{Joint Quantum Institute, National Institute of Standards and Technology and University of Maryland, 
Gaithersburg, Maryland 20899, USA}

\date{\today}
\begin{abstract}

We load cold atoms into an optical lattice dramatically reshaped by radiofrequency (rf) coupling of state-dependent lattice potentials.   This rf dressing changes the unit cell of the lattice at a subwavelength scale, such that its curvature and topology departs strongly from that of a simple sinusoidal lattice potential.  Radiofrequency dressing has previously been performed at length scales from mm to tens of $\mu$m, but not at the single-optical-wavelength scale.    At this length scale significant coupling between adiabatic potentials leads to nonadiabatic transitions, which we measure as a function of lattice depth and dressing frequency and amplitude.   We also investigate the dressing by measuring changes in the momentum distribution of the dressed states.   

\end{abstract}
\maketitle

For almost twenty years, optical lattices have been used to cool and confine neutral atoms, recently leading to condensed-matter-analog systems,  optical clocks and possible platforms for quantum computing~\cite{PhysRevLett.81.3108,Greiner:2002fk,PhysRevLett.82.1975,PhysRevLett.82.1060,Anderlini:2007qy,Takamoto:2005lr}.      The appeal of optical lattices arises from their versatility and tunability, e.g., through control of the topology, depth, and spin dependence of the lattice~\cite{PhysRevLett.91.010407,PhysRevLett.87.160405,lee:020402,sebby-strabley:200405}.   Here we extend this control by rf dressing a spin-dependent lattice, a technique which has been recently proposed~\cite{daleypreprint}.   The dressed lattice comprises unit cells whose localized adiabatic eigenstates are spatially-varying superpositions of the bare spin states.   The adiabatic potentials have new subwavelength structure, which in certain limits can be ring-like within a unit cell.  

The creation of rf-dressed adiabatic potentials in inhomogeneous magnetic fields is well-known experimentally,  from the use of rf transitions in evaporative cooling~\cite{ketterle1996ect}  to the creation of shell-like quasi-two-dimensional systems in magnetically-trapped condensates~\cite{PhysRevLett.86.1195,white:023616,perrinbec}, and the use of rf dressing to effect condensate splitting and interferometry in micron-scale chip traps~\cite{Schumm:2005lr}.      Optical lattices with sub-half-wavelength structure in 1D have been generated by several groups through the use of multiphoton Raman processes ~\cite{ritt:063622,PhysRevLett.76.4689,zhang:043409} or static magnetic field couplings~\cite{PhysRevLett.85.3365}.  The approach we present here, using the rf dressing of a state-dependent lattice, creates 2D subwavelength structure beyond simple lattice period division.   We create such a dressed lattice, and load ultracold atoms into the ground band of the uppermost adiabatic potential.

\begin{figure}[hb]
\centering
  \includegraphics[width=\columnwidth]{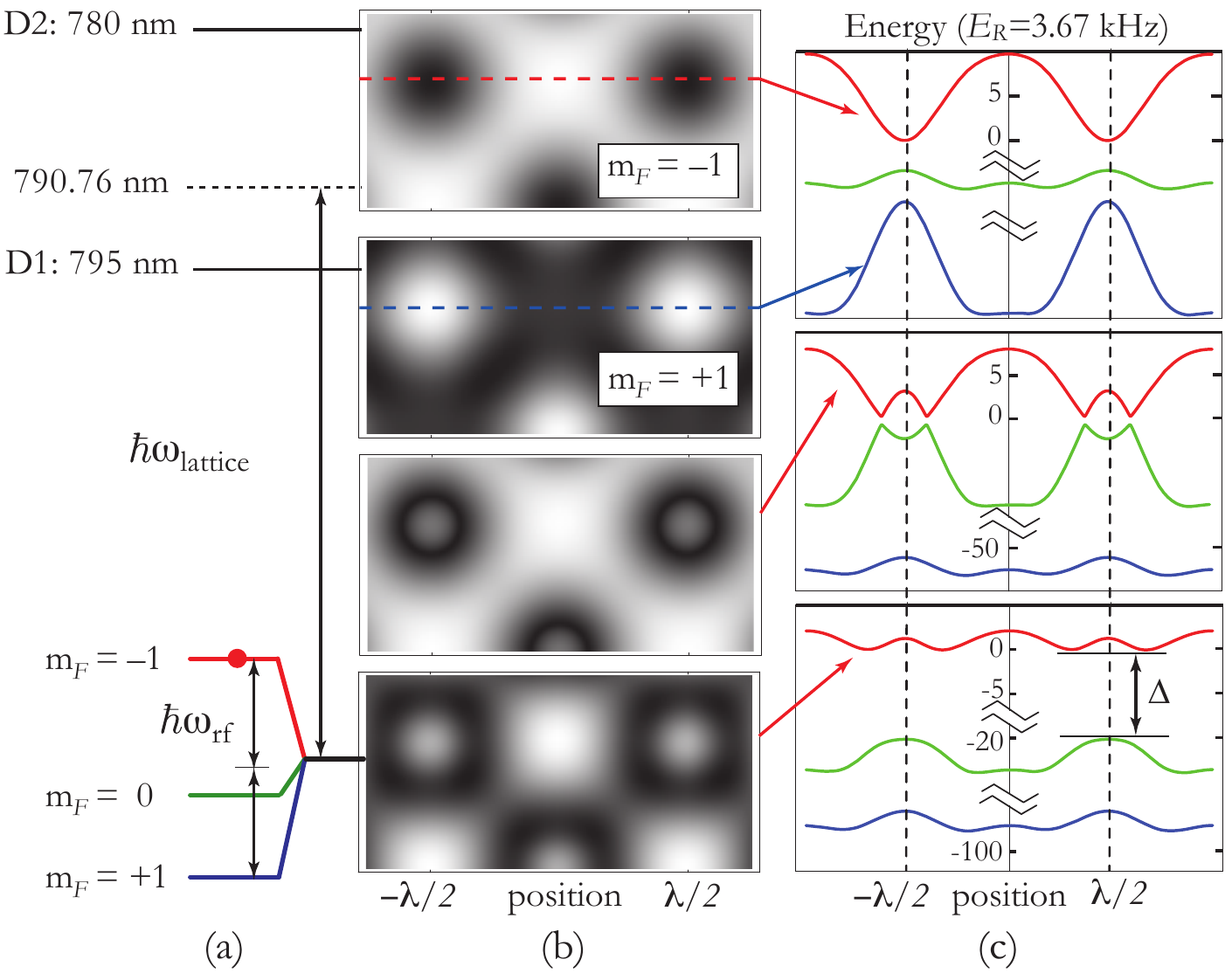}   
    \caption{(color online) (a) The energy levels of $^{87}$Rb relevant to our lattice; the rf coupling can selectively address either $m_F$ transition due to the quadratic Zeeman shift. (b)  Top two images: the bare lattice potentials for $m_F=\pm 1$, exhibiting approximately opposite light shifts.  Bottom two images: the uppermost adiabatic potential, weakly dressed and strongly dressed.  (c)  From top: bare lattice potentials for $U\simeq10E_R$; dressed adiabatic potentials at $\omega_{\rm rf}/2\pi=35.90$ MHz, $\Omega/2\pi=20$ kHz; dressed adiabatic potentials at $\omega_{\rm rf}/2\pi=35.90$ MHz, $\Omega/2\pi=205$ kHz.   Note the significantly increased gap for stronger coupling.}    
\label{geometry}
\end{figure}
%------------------

For these state-dependent lattice experiments we use a spin-polarized $^{87}$Rb ($5^2S_{1/2}$ $|F=1,m_F=-1\rangle$) Bose-Einstein condensate (BEC).    The state dependence is produced by tuning the lattice laser between the $^{87}$Rb D1 and D2 resonances, where vector light shifts are important~\cite{lee:020402}.    Applying an rf field  of appropriate amplitude and frequency creates an adiabatic potential whose shape is distorted from the bare potentials, as shown in Fig.~\ref{geometry}(b). 
After holding the atoms in the dressed lattice for a variable time, we abruptly turn off the lattice and observe the momentum distribution by imaging the atoms after time-of-flight (TOF) time $t_{\rm TOF}$.    This gives information about the structure of the dressed lattice, as well as a measure of the lifetime of the dressed adiabatic eigenstates.

Our apparatus, described elsewhere~\cite{sebby-strabley:033605},  produces BECs of $\sim$$10^5$ atoms in a Ioffe-Pritchard magnetic trap.   We load the $m_F=-1$ condensate into the ground band of a 3D optical lattice via an exponentially increasing intensity ramp of duration 300 $\mu$s ($\tau=$ 50 $\mu$s), a timescale adiabatic with respect to band (vibrational) excitation.  We then remove the magnetic trapping fields, leaving a uniform bias field of 5.117(3) mT
\footnote{Unless otherwise stated, all uncertainties herein reflect the uncorrelated combination of 1$\sigma$ statistical and systematic uncertainties.}, 
 a Zeeman resonance $\nu_{-1,0}= 36.12(2)$ MHz and a quadratic Zeeman shift $\delta^\prime/2\pi=\nu_{-1,0} -\nu_{0,+1} =$   376(1) kHz, where $h \nu_{m,m^\prime}$ is the positive energy difference between Zeeman sublevels $|F=1,m_F\rangle$.

The lattice in the $\hat{x}$-$\hat{y}$ plane ($xy$ lattice) is similar to that described in~\cite{sebby-strabley:033605}.   We tune a Ti:sapphire laser to $\lambda=$790.76~nm, where for $\sigma^+$ light, atoms in  the three $F=1$ Zeeman sublevels $m_F=$$-$1, 0, +1 experience total light shifts in the approximate ratio of $-$3 :   1 : 5.    This  can be understood as the sum of a state-independent scalar light shift and a state-dependent vector light shift---an effective magnetic field related to the local ellipticity of the optical polarization.    For appropriate phase shifts of the lattice beams~\cite{lee:020402}Œ, the combination of the light shifts and the bias field along $(\hat{x}-\hat{y})/\sqrt{2}$ results in the potentials shown in Fig.~\ref{geometry}.

The atoms are confined along $\hat{z}$ (the direction of gravity) using a 1D optical lattice, derived from the same Ti:sapphire laser~\footnote{The vertical lattice beams deviate from counterpropagation by 18$^\circ$, and also differ in frequency from the 2D lattice beams by $\simeq$ 160 MHz.}.     These lattice beams are linearly polarized, yielding an effectively blue-detuned and spin-independent lattice.   As measured by pulsed-lattice diffraction~\cite{PhysRevLett.83.284} this vertical lattice is 6$E_R$ deep, where $E_R=\hbar^2 k^2 /2M \simeq h\times 3.68$ kHz, $k=2\pi/\lambda$ and $M$ is the atomic mass.   This lattice, together with the  $xy$ lattice, fully confines the atoms in 3D, and we do not observe any displacement due to gravity over the duration of the experiment, even though the magnetic trap has been removed.    

%------------------
\begin{figure}[t]
\begin{center}
 \includegraphics[width=\columnwidth]{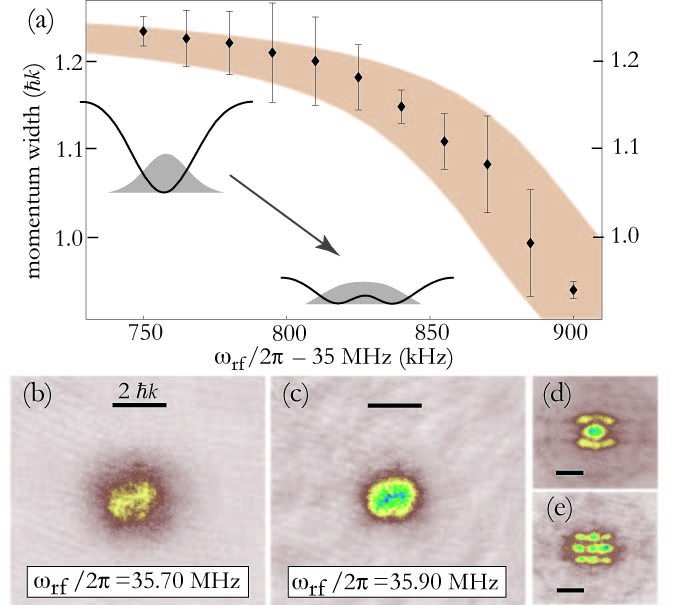}   
\caption{(color online) (a) The width of the momentum distribution as a function of final dressing frequency, as measured by absorption imaging.    The width is calculated as $Md/\hbar t_{\rm TOF}$, where $d$ is the Gaussian $1/e$ radius along one direction, corrected for the initial cloud size.    The displayed uncertainties represent  1$\sigma$ shot-to-shot scatter of five points.   The shaded area represents predicted widths given the values and estimated uncertainties of the magnetic field (5.117(3) mT) and the depth of the bare lattice $U=10.0(5) E_R$.    Also shown are the bare and dressed single-site potentials (solid line) and ground-state wavefunctions (gray).  (b,c)  Sample images of the off-resonant (effectively bare) and dressed momentum distributions respectively (both at $\Omega/2\pi=$ 205 kHz) showing significant narrowing of the latter.  (d,e) Examples of dressed momentum distributions created through faster ($>$ 1 MHz/ms) rf ramps, such that higher vibrational states of the dressed lattice are populated.}
\label{width}
\end{center}
\end{figure}

After loading, we wait 30 ms for the atoms to dephase and fill the lowest Bloch band of the lattice~\cite{PhysRevLett.87.160405}, such that a sudden-release TOF image will approximate the single-lattice-site momentum distribution (with resolution limited mainly by the spatial extent of the atom cloud).   The dephasing time is well within the $1/e$ lifetime of atoms in the lattice due to spin-flip loss from light scattering, which we measured to be 85 ms at an $xy$ lattice depth of $U=55 E_R$ for $m_F=-1$ atoms; this lattice depth is more than twice the largest used in the experiment.      The depth of the $xy$ lattice was measured with two-photon Raman vibrational spectroscopy~\cite{perrin1998scn}.

We dress the lattice with an rf magnetic field aligned perpendicular to the bias field.     The single-particle rotating-frame Hamiltonian, taking the rotating-wave approximation (RWA), is $\mathcal{H} ={\bf p}^2/2M+{\mathcal H}_1({\bf r})$ , where

\begin{equation}
\mathcal{H}_1({\bf r})=\left(\begin{array}{ccc}V_{-1}({\bf r})-\delta & \Omega/2 & 0 \\ \Omega/2 & V_{0}({\bf r}) & \Omega/2 \\0 & \Omega/2 & V_{+1}({\bf r})+\delta+\delta^\prime\end{array}\right)
\end{equation}
and $\delta = \omega_{\bf rf}- 2\pi\nu_{-1,0}$ .  The state-dependent lattice potentials $V_{m_F}({\bf r})$ (see Fig.~1) are calculated using our model of the lattice~\cite{sebby-strabley:033605}.    $\Omega$ is obtained through observations of rf-driven oscillations between $m_F=-1$ and $m_F=0$ (in a spin-independent lattice) at rates up to a maximum of $\omega_{\rm osc}/2 \pi=$ 200 kHz.   $\Omega$ differs slightly from $\omega_{\rm osc}$ due to the multilevel nature of this system.    Diagonalizing only $\mathcal{H}_1$ (i.e., the Born-Oppenheimer approximation) gives rise to adiabatic potentials;  failure of this approximation will manifest as momentum-dependent couplings between the adiabatic potentials.   

We apply rf beginning  well below resonance($\simeq-1$ MHz), sweeping at constant rate (300 kHz/ms) over $\simeq$ 3 ms to a final near-resonant  frequency, and hold the newly created dressed state for a variable time.   The lattice beams, rf field, and bias field are then turned off in $\lesssim$ 1 $\mu$s, $\simeq$ 10 $\mu$s, and $\simeq$ 300 $\mu$s, respectively.  After TOF of 12.2 ms we observe the atoms with resonant absorption imaging along the $\hat{z}$ direction.    

Fig.~\ref{width}(a) depicts the observed width of the momentum distribution (from a Gaussian fit to the TOF density distribution)  as a function of final frequency at maximum rf power.   The optical power in the $xy$ lattice was fixed such that the bare depth $U$ was $\simeq 10E_R$.   Also shown is the expected width from the calculated ground state momentum wavefunction in the adiabatic potential.   The data show narrowing of the momentum distribution as the lattice is dressed, implying significant alteration of the wavefunction and hence the lattice structure itself.   At these dressed depths, while the potential exhibits ringlike character, the ground state is still simply connected: see Fig. 2(a). The observed dressed state (corresponding to $m_F=-1$ without rf) is predominantly a superposition of $m_F=\pm1$ at these rf detunings~\footnote{Separation of the dressed clouds using a magnetic field gradient during TOF revealed the expected spin character.}.      Unlike in a two-level system, the avoided-crossing energy gap  $\hbar\Delta$ between adiabatic potentials is not simply related to $\Omega$.   For $\Omega \ll \delta^\prime$, $\Delta\simeq\Omega^2/\delta^\prime$, and for $\Omega \gg \delta^\prime$, $\Delta \simeq \Omega/\sqrt{2}-\delta^\prime/4$.      For  $\Omega/2\pi=$ 205 kHz, the largest used in this experiment, and $\omega_{\rm rf}=$ 35.90 MHz, we calculate  $\Delta/2\pi\simeq$ 75 kHz, a value only weakly dependent on $U$.    To further characterize our dressed lattice, we performed Raman vibrational spectroscopy, but observed that the Raman signal did not persist beyond weak dressing.   This may be due to to shrinking matrix elements between the combined spin and spatial eigenfunctions of the dressed bands, or increasing width of the excited bands of the (shallower) dressed lattice. 

Three adiabaticity criteria are relevant to loading and holding atoms in the ground band of the dressed lattice:   (1) Adiabatic following of the local spin eigenstates of ${\mathcal H}_1({\bf r})$ during the upward rf sweep $\delta(t)$  is well-satisfied.   (2) Deformation of the dressed lattice with respect to vibrational excitation is nearly adiabatic, yielding predominantly ground-band occupation.    Faster sweep rates yielded interesting deviations from this condition, shown in Fig.~\ref{width}(d,e), in which we observe lobed structure in the momentum distributions indicating higher vibrational states.     (3) The Born-Oppenheimer approximation, i.e., the degree to which zero-point motion in the lattice does not induce transitions between adiabatic potentials, is more difficult to satisfy.     In previous experiments involving rf-dressed adiabatic potentials, the length scales were large enough such that this condition was easily met for cold atoms~\cite{white:023616,perrinbec,Schumm:2005lr,PhysRevLett.86.1195}.   For our experiment, the bare lattice confinement is such that, even for our largest rf coupling, nonadiabatic loss is a factor; in our experiment these loss rates are $10^1$--$10^3$ s$^{-1}$.    These losses do play a role in the loading process itself, but this is irrelevant as long as enough atoms are loaded such that subsequent loss is measurable.

%------------------
\begin{figure}[t]
\begin{center}
\includegraphics[width=\columnwidth]{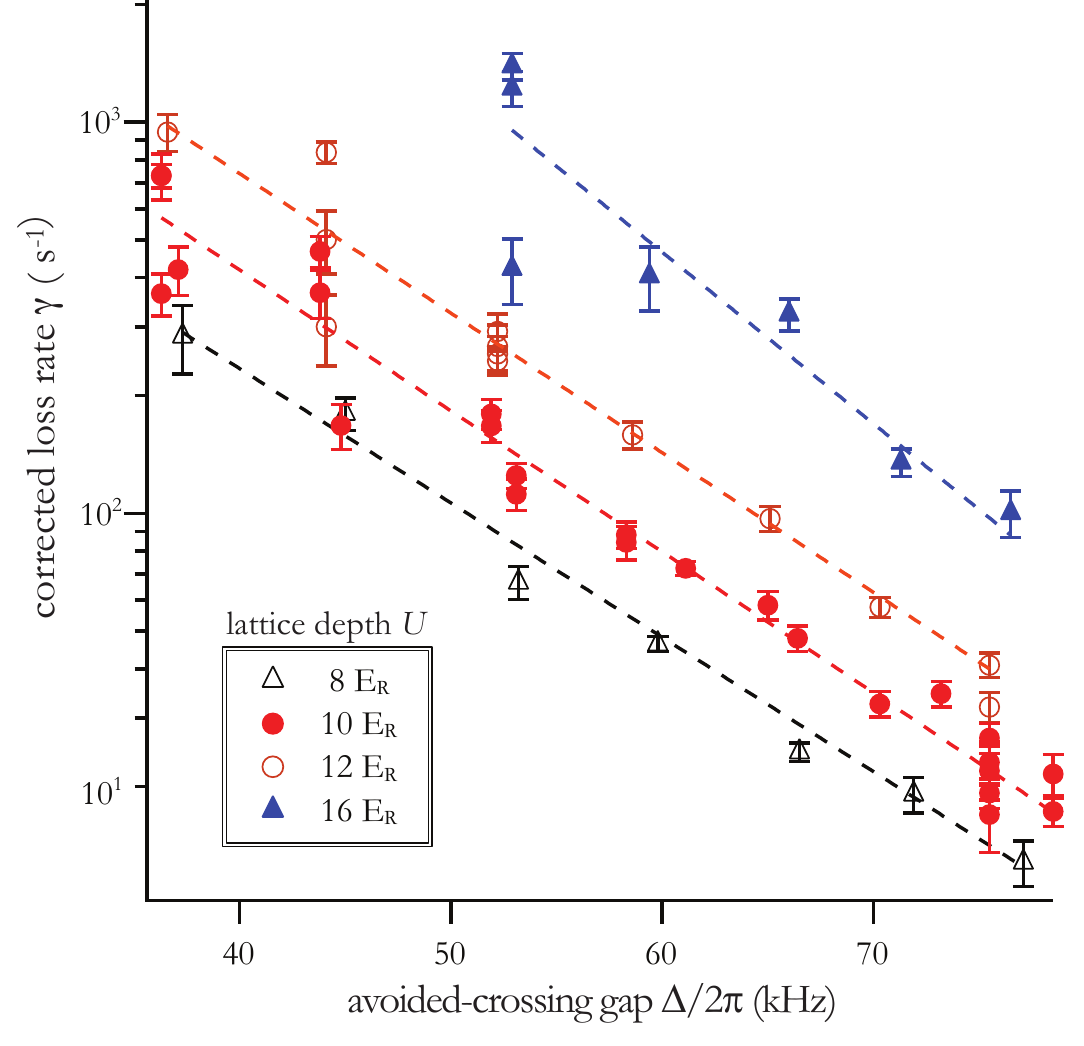} 
\caption{(color online) Dressed-state losses as a function of calculated avoided-crossing gap $\Delta/2\pi$, for four bare lattice depths.   Vertical bars represent uncertainties-of-fit; multiple points near a given $\Delta$ indicate scatter in multiple runs.    The data are well-represented by $\gamma(\Delta)=A e^{-B\Delta/2\pi}$, with $B=$ 83(6) $\mu$s , 83(11) $\mu$s, 82(13) $\mu$s , and 100(30) $\mu$s, for $U=8 E_R$, $10 E_R$, $12 E_R$, and $16 E_R$ respectively.  The scaling of the data does not change if plotted versus $\Omega$.}
\label{lossvsrabi}
\end{center}
\end{figure}
%------------------

\begin{figure}[t]
\includegraphics[width=\columnwidth]{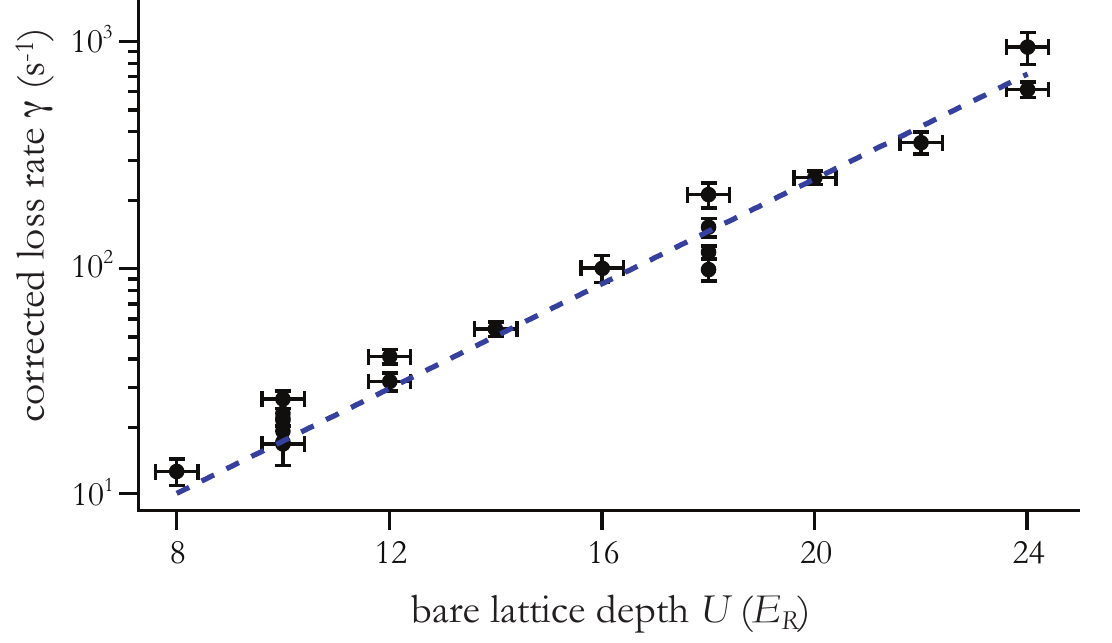} 
\caption{(color online) Dressed-state losses as a function of bare lattice depth $U$, at constant  $\Omega/2\pi=205$ kHz and dressing frequency 35.900(35.875) MHz for the first five (last four) depths, resulting in approximately constant gap $\Delta/2\pi\simeq 75$ kHz.   Vertical bars represent uncertainties-of-fit; multiple points at a given depth represent the scatter in multiple runs.   Horizontal bars represent known drifts in laser power; not shown is a constant systematic uncertainty in depth determination on the order of 5\%.    The data are well-represented by a simple exponential fit $\gamma(U)=Ce^{D U}$, yielding $C=1.2(4) {\rm s}^{-1}$ and $D=0.27(3) E_R^{-1}$.   }   
\label{lossvsoptical}
\end{figure}

We load atoms into the uppermost adiabatic dressed potential using the procedure described above, and measure the number of atoms remaining in the central momentum feature, as in Fig. 2(b-e), as a function of hold time.   Atoms transitioning to lower adiabatic potentials appear in absorption images as high-momentum rings of $\sim10\hbar k$ (well outside the range of Fig. 2(b-e)), and are counted as loss.    Fig.~\ref{lossvsrabi} shows loss rates as a function of calculated avoided-crossing gap $\Delta$ as a function of $U$, at dressing frequencies near 35.90 MHz.      The rates are first obtained through two-parameter fits to an exponential decay and then slightly corrected via subtraction of the appropriately scaled spin-flip loss rate ($1/85~{\rm ms}^{-1}$ for $55E_R$).    The data are well-described by $\gamma(\Delta)=Ae^{-B\Delta}$.   Figure~\ref{lossvsoptical} shows loss rate at constant $\Omega$ as a function of $U$, well described by $\gamma(U)=Ce^{DU}$.

We consider our observations in the context of semiclassical theory for the losses. The traversal of a two-level avoided crossing is often described by Landau-Zener (L-Z) theory, which for a particle traveling along $x$ at velocity $v$ through an avoided crossing characterized by a gap $\Delta$ and bare energy levels whose difference $E(x) = E x$, the adiabatic following probability is $P_A=1-{\rm e}^{-\frac{1}{2} \pi \hbar \Delta^2/(v E')}$.   A semiclassical scaling argument for the motion of an atom in the ground state of a lattice site of depth $U$ yields $v \propto U^{1/4}$ and $vE^\prime \propto U^{5/4}$, and a loss rate scaling as $\gamma\propto\omega_{l} (1-P_{\rm A})$, where $\omega_l$ is a trapping frequency and $(1-P_A)\ll 1$.   For a particle of insufficiently large energy with respect to $\Delta$ (i.e., atoms trapped in the ground state of an adiabatic potential), the simple L-Z  approach may not apply.   In a more sophisticated semiclassical approximation~\cite{kazantsev1990mal,daleypriv}, more appropriate to ground-state trapped atoms, the resulting transition rates can be estimated as $\gamma \propto \omega_l {\rm e}^{-\alpha \Delta/\omega_l}$, where $\omega_l \propto \sqrt{U/\Delta}$ is the trapping frequency ($\omega_l \ll \Delta$) of the uppermost adiabatic potential, and $\alpha\sim 1$ is a constant. The shape of our adiabatic potentials differs significantly from this simple case, but the scaling suggests transition rates that change exponentially with the coupling gap and bare lattice depth.

%------------------
%------------------.   

Unsurprisingly, the L-Z model does not describe the data in Figs.~\ref{lossvsrabi} and \ref{lossvsoptical};  the exponential coefficient scales as $-\Delta$ rather than $-\Delta^2$ and also scales as $U$, not $-1/U^{5/4}$.    Fits to the L-Z model are significantly worse than those in Figs. 3 and 4, although they agree  with the general trends.    Similar results were found upon calculating a Fourier-weighted sum of the L-Z formula over the dressed 2D ground-state wavefunction, $\int |\psi_{\bf k}|^2 P_{\rm NA}({\bf k}) d{\bf k}$.     The more sophisticated semiclassical approximation performs better than the L-Z approach, but still fails to satisfactorily predict the observed scaling.    It is likely that 1D semiclassical descriptions are simply inadequate for this 2D quantum problem; perhaps a more complete solution will yield the very simple scaling we observe.   Furthermore, heating to higher vibrational levels of the uppermost adiabatic potential, likely caused by instability in the rf dressing frequency (measured to be on the order of several kHz), is likely to increase the nonadiabatic loss rate due to increased mean velocity in higher bands.   

While the dressed potentials shown in Fig.~\ref{geometry}(b) are ringlike, the corresponding ground-state wavefunctions are generally not.     Nevertheless, ringlike wavefunctions with reasonable lifetimes would be possible had we sufficient rf coupling; based on the data presented here, we extrapolate that ringlike wavefunctions of lifetime $\simeq 100$ ms should appear for a coupling $\Omega/2\pi\simeq 400$ kHz and  $U\simeq55 E_R$.      

This paper presents observations of a novel lattice potential consisting of rf-coupled components of a state-dependent optical lattice.   The lifetime of this dressed lattice depends exponentially on both coupling strength and optical lattice depth.   We observe lifetimes up to 100 ms for strongly-dressed shallow lattices, a timescale sufficient for the realization of many-body condensed-matter systems.    Extensions of the band-structure engineering demonstrated by this experiment lead to lattice geometries with structures much more complex than possible with conventional optical lattices, leading to intriguing single-site wavefunctions and the next-nearest-neighbor interactions required for extended Bose-Hubbard models~\cite{scarola:051601}. 

\begin{acknowledgments}
We thank P.~Zoller for inspiring this particular experiment and A.~J.~Daley, G.~Pupillo, and P.~Z. for ongoing helpful discussions.  This work was partially supported by DTO and  ONR.  N.L., P.J.L., and B.L.B. acknowledge support from NRC fellowships. 
\end{acknowledgments}

%\bibliographystyle{apsrevNOURL}
%\bibliography{master}

\end{document}